\documentclass[aps,prb,superscriptaddress,showpacs,floatfix,preprint]{revtex4}
\usepackage{graphicx}
\usepackage{setspace}
\usepackage{rotating}

\def\qcut{{q_0^{\rm cut}}}
\def\eg{{\it e.g.}}
\def\ie{{\it i.e.}}
\def\cf{{\it c.f.}}
\begin{document}

\title{Van der Waals density functionals applied to solids}
\author{Ji\v{r}\'\i\ Klime\v{s}}
\affiliation{London Centre for Nanotechnology and Department of Chemistry, University College London, London, WC1E 6BT, UK}
\author{David R. Bowler}
\affiliation{London Centre for Nanotechnology and Department of Physics and Astronomy, University College London, London, WC1E 6BT, UK}
\author{Angelos Michaelides}
\email{angelos.michaelides@ucl.ac.uk}
\affiliation{London Centre for Nanotechnology and Department of Chemistry, University College London, London, WC1E 6BT, UK}

\pacs{
71.15.Mb 
61.50.-f 
31.15.-p 
}

\date{\today}

\begin{abstract}

The van der Waals density functional (vdW-DF) of Dion~{\it et al.} [Phys. Rev. Lett. {\bf 92}, 246401 (2004)]
is a promising approach for including dispersion in approximate 
density functional theory exchange-correlation functionals. 
Indeed, an improved description of systems held by dispersion forces has been demonstrated in the literature.
However, despite many applications, standard general tests on a broad range of materials are lacking.
Here we calculate the lattice constants, bulk moduli, and atomization energies for a range of solids using 
the original vdW-DF and several of its offspring.
We find that the original vdW-DF overestimates lattice constants in a similar manner to how it overestimates
binding distances for gas phase dimers. 
However, some of the modified vdW functionals lead to average errors which are similar to those of PBE or better.
Likewise, atomization energies that are slightly better than from PBE are obtained from the modified vdW-DFs.
Although the tests reported here are for ``hard" solids, not normally materials for which dispersion forces
are thought to be important, we find a systematic improvement in cohesive properties for the alkali metals and alkali halides
when non-local correlations are accounted for.

\end{abstract}

\maketitle


\section{Introduction}


London dispersion interactions are an ubiquitous phenomenon which contribute to the stability of a wide
variety of systems ranging from biomolecules to molecules adsorbed on surfaces. 
However, the origin of the dispersion forces -- non-local electron-electron correlations -- makes
their accurate theoretical description challenging.
This is especially true for density functional theory (DFT)
where local or semi-local functionals lack the necessary ingredients 
to describe the non-local effects. 
In fact, developing methods that include dispersion, at least approximately, 
has been one of the most important fields of development in DFT in the last decade.
Out of the various schemes that have been proposed to add dispersion to current DFT 
approximations~\cite{andersson1996,dobson1996,dion2004,lilienfeld2004,antony2006,becke2007,grimme2007,sato2007,tkatchenko2009,sato2010,sato2010sec},
the van der Waals density functional (vdW-DF) method~\cite{dion2004} is very appealing 
since it is based directly on the electron density.
In this functional the exchange-correlation energy takes the form of
\begin{equation}
E_{\rm xc}=E_{\rm x}^{\rm GGA}+E_{\rm c}^{\rm LDA}+E_{\rm c}^{\rm nl}\,,
\end{equation}
where the exchange energy $E_{\rm x}^{\rm GGA}$ uses the revPBE generalized-gradient approximation
 (GGA) functional~\cite{zhang1998}, 
$E_{\rm c}^{\rm LDA}$ is the local density approximation (LDA) to the correlation energy. 
The $E_{\rm c}^{\rm nl}$ is the non-local energy term
which accounts approximately for the non-local electron correlation effects. 
Although $E_{\rm c}^{\rm nl}$ is obtained using a relatively simple double space integration, 
this still represents an improvement compared to local or semi-local functionals.

Although the vdW-DF method greatly improves the interaction energies of dispersion bonded systems, 
its accuracy, however, has been shown~\cite{gulans2009,kelkkanen2009,klimes2010} to be inferior to certain GGAs 
for a range of systems where hydrogen bonds are present~\cite{jurecka2006, li2008, santra2008}.
This has lead to modifications of the method that have focused on both the exchange and correlation 
parts~\cite{vydrov2009prl,klimes2010,vydrov2010pra,lee2010,cooper2010,vydrov2010better}. 
With several functional forms proposed, it is important to test the methods on general reference test sets
to uncover strengths and weaknesses and help further development.
To this end we assess here the functionals using a test set of solid state properties of materials~\cite{staroverov2004}.
Apart from method assessment, this is also important since many of the applications 
of the vdW functionals
lie outside of ``soft matter", involving, for example, adsorbates on solid surfaces.
Indeed there has recently been a surge of interest in the application of vdW-DF to adsorption processes, 
including adsorption of water or of 
hydrocarbons on different surfaces~\cite{lazic2005,chakarova2006,sony2007,johnston2008,yanagisawa2008,moses2009,berland2009,chen2010,carrasco2011}.
Such studies will require a good description of the structural parameters of the substrate
to accurately describe the systems of interest~\cite{feibelman2008}.
The PBE lattice constant is usually employed in studies using revPBE-vdW, however, as we show here,
the PBE and revPBE-vdW lattice constants can differ by several percent and the surface is therefore
artificially strained.

There is at least one more reason to perform these tests: non-local correlations are thought to be
important for solid state materials where the core electron densities have relatively large polarizability.
For example, copper and gold have been
subject to several studies~\cite{friedel1952,rehr1975,richardson1977},
with the vdW contribution to binding estimated to be 0.2--0.6~eV for
Cu and 0.6--1.2~eV for Au (\ie, up to $\sim$30\% of the total atomization energy of bulk Au). 
There are only a handful of studies concerning heavy alkali
metals (see, \eg, Ref.~\onlinecite{upadhyaya1979}), but it was found that dispersion needs to be included to make
the bcc structure the most stable.
More recently, the need to include non-local correlation in DFT semi-local functionals
has been discussed for the heavy alkalis~\cite{tao2010}.

Here we test the performance of several vdW functionals using a standard test of
lattice constants, bulk moduli, and atomization energies of solids. 
Our test is similar to the test of Csonka~{\it et al.}~\cite{csonka2009} and includes
metals, ionic, and covalent materials.
We include the original vdW-DF (referred to as revPBE-vdW herein), 
the recently proposed vdW-DF2~\cite{murray2009,lee2010} (referred to as rPW86-vdW2 herein), 
and two vdW functionals developed recently by us, optPBE-vdW and optB88-vdW~\cite{klimes2010}.
We also propose a new exchange functional based on the B86b exchange~\cite{becke1986b} which gives 
an accuracy similar to the optB88 based vdW-DF on the S22 reference set of weakly bonded gas phase 
dimers~\cite{jurecka2006} and has an improved asymptotic 
behavior~\cite{murray2009}. See Appendix~\ref{app_optb86b} for more details of the optB86b exchange functional.
This test, together with our previous study~\cite{klimes2010}, should give the reader a broad overview of the
strengths and deficiencies of the vdW functionals that will hopefully lead to further developments.

The main outcomes of this study are: 
the revPBE-vdW and rPW86-vdW2 functionals significantly overestimate the lattice constants for most materials
considered and the average absolute error is more than twice that of the optB88-vdW and optB86b-vdW functionals.
The optB88-vdW and optB86b-vdW functionals give errors between that of PBEsol and PBE
with optB86b-vdW giving smaller errors than optB88-vdW.
This is because the exchange enhancement factor ($F_x$) of the optB88-vdW 
and optB86b-vdW functionals is between the $F_x$ of PBEsol and PBE for small reduced density gradients ($s$).
The optB88-vdW and optB86b-vdW  functionals also almost halve the errors of PBE in atomization energies.
optPBE-vdW improves over revPBE-vdW but not as much as optB88-vdW does.
This behavior for the lattice constants is similar to that of binding curves and bond lengths;
in all three cases the functionals with rapidly growing enhancement factors give on average longer
equilibrium distances and agree less with the reference values than the functionals
where $F_x$ follows the slowly varying gas limit for small $s$.

In the next section we discuss the implementation of the vdW-DF method and details of the 
computational setup. 
The results are summarized in Sections~\ref{secIII}, \ref{secIV}, and \ref{secV} for lattice constants, 
bulk moduli, and atomization energies, respectively. We study the differences between local, 
semi-local, and non-local correlation functionals in Section~\ref{secVI}. 
And finally in Section~\ref{secVII} we discuss the implications of this study for further development of the vdW-DF methods.

\section{Computational setup}

We have used the VASP~\cite{kresse1993,kresse1996} code with our implementation of the vdW-DF correlation
using the efficient algorithm of Rom\'an-P\'erez and Soler~\cite{soler2009}.
The vdW-DF term is calculated on the sum of the pseudo valence and partial electronic core charge densities,
\ie, on the same density that is used to calculate the valence exchange-correlation 
energy in the projector-augmented wave (PAW)~\cite{blochl1994,kresse1999} method in VASP.
The use of the PAW method means that the calculation is all-electron frozen core (with PBE orbitals)
for the exchange and the LDA correlation part of the exchange-correlation energy. 
The evaluation of the vdW correlation energy is done in a pseudopotential approximation.
We test this approximation in Appendices B and C and find that it is very accurate.
For example, the error in lattice constant is usually below 0.1\%, slightly higher for materials with 
very small bulk moduli.
Such differences are much smaller than the intrinsic errors
of the exchange-correlation functionals themselves and smaller or comparable to differences between different
codes~\cite{paier2006,haas2009,csonka2009,harl2010} or potentials~\cite{mattsson2008}.
While to obtain high accuracy within the Rom\'an-P\'erez and Soler scheme using the all-electron density
a careful choice of parameters is required (discussed in Appendices B and C), the VASP calculations 
are accurate using less tight settings.
For VASP calculations we use 30 interpolation points for the $q_0$ function with a saturation value $q_0^{\rm cut}=10$.
The vdW kernel uses a hard setting for the kernel short range softening which eliminates the need for the
soft correction term (see Ref.~\onlinecite{soler2009}). 
Finally, we note that the algorithm utilizes the fine FFT grid and, except for the FFTs and summing of the energy, 
there is no other communication needed between the processes. 

We employ a standard approach to calculate the solid properties. The energy is calculated for a set
of lattice constant values and for each functional at least seven points around the lowest energy 
are used to fit the Murnaghan equation of state. 
The plane-wave basis cut-off is set to 750~eV (900~eV for solids containing C or F).
To reduce errors, we have used the latest hard PAW potentials supplied with VASP~\cite{paier2005}
with the highest number of valence electrons.
For semiconductors and ionic 
solids (metals) a 8$\times$8$\times$8 (16$\times$16$\times$16) Monkhorst-Pack k-point grid is used
in the conventional unit cell. Our PBE lattice constants agree well with the VASP calculations 
of Paier~{\it et al.}~\cite{paier2006} as well as with the all-electron reference PBE and PBEsol values 
of Haas~{\it et al.}~\cite{haas2009}. 
The reference calculations for atoms were performed in a large $12\times14\times16$~\AA$^3$ box;
for spin polarized atoms we evaluate the $E_{\rm c}^{\rm nl}$ term on the sum of the two spin-densities.
The experimental reference
values, corrected for zero point energy effects in the case of lattice constants and atomization energies, 
are taken from Refs.~\onlinecite{csonka2009} and~\onlinecite{perdew2009rev}.
The statistical values that we use to quantify the errors of the functionals are 
the mean error (ME) and the mean absolute error (MAE), as well as the relative versions of these 
quantities, namely mean relative error (MRE) and mean absolute relative error (MARE).

\section{Lattice Constants}
\label{secIII}

\begin{figure}[h!]
\centerline{
\includegraphics[height=9cm,angle=-90]{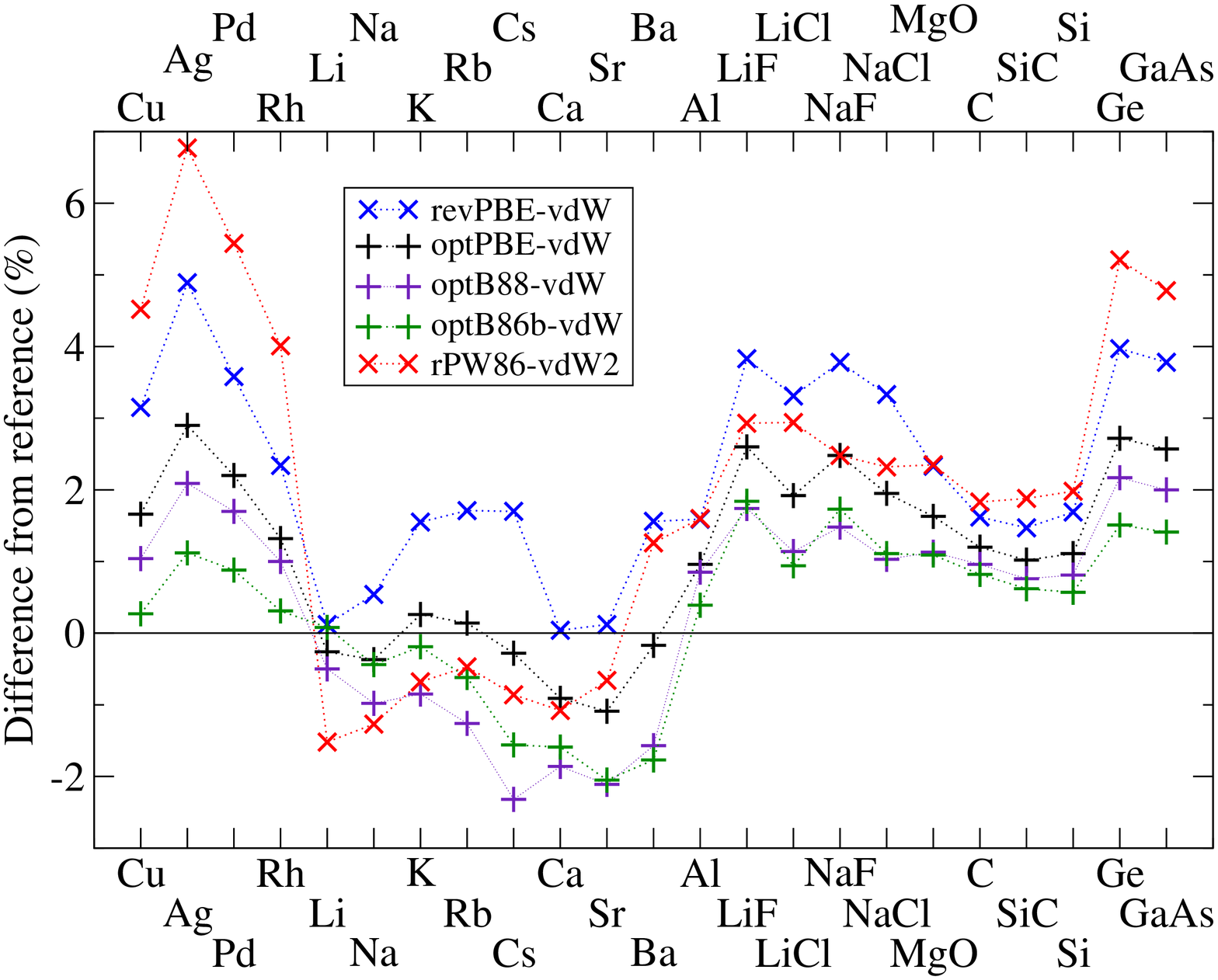} 
\includegraphics[height=9cm,angle=-90]{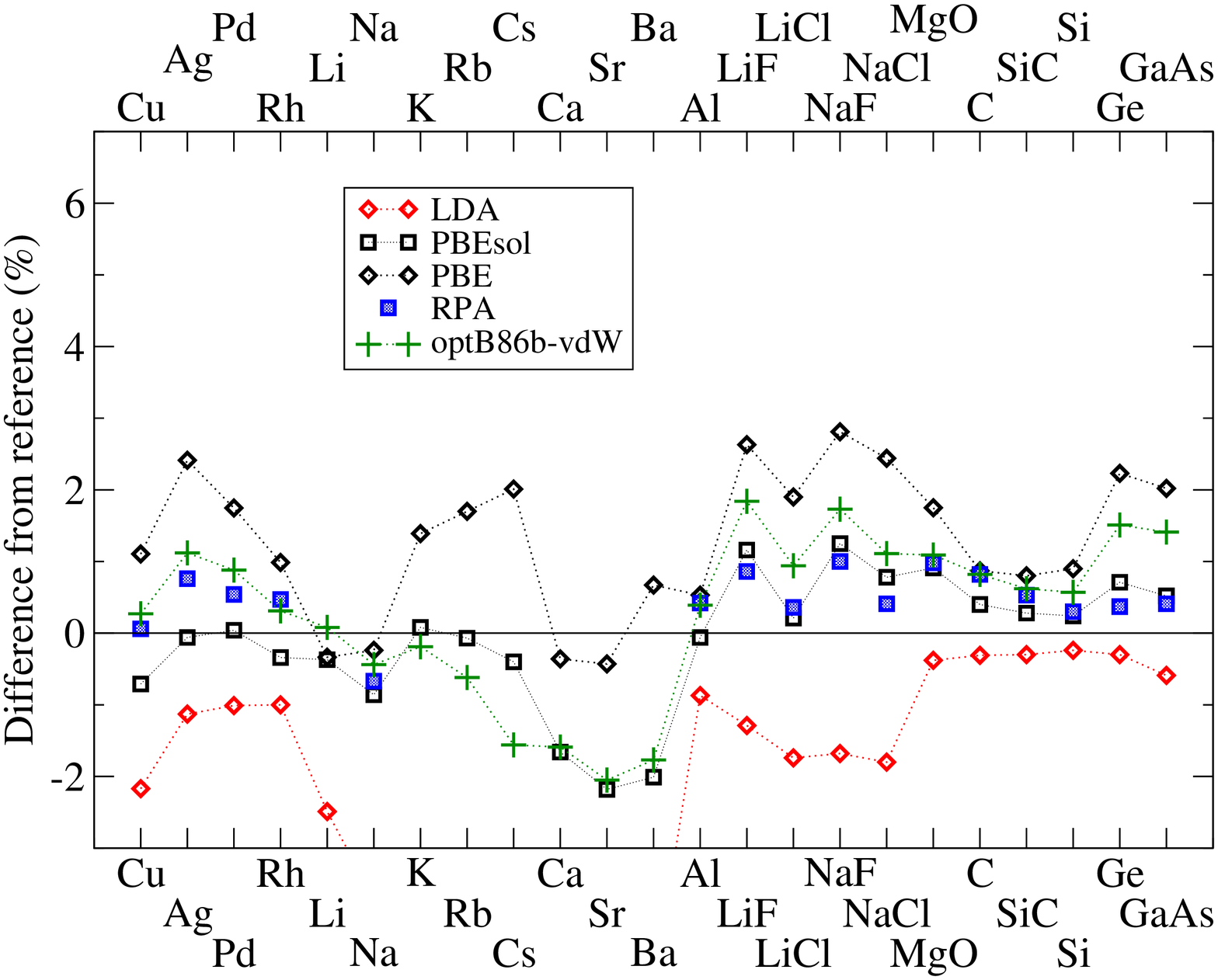}
}
\caption[Relative erros in lattice constants]
{Comparison of the relative errors in the lattice constants calculated with different vdW-DF functionals, 
LDA (from Ref.~\onlinecite{csonka2009}), PBE, PBEsol, and recent data using the RPA~\cite{harl2010}. 
The vdW methods are shown in the left panel while the right panel contains the results of the (semi-)local functionals, 
RPA as well as the optB86b-vdW results for comparison.
All the methods overestimate the lattice constants for 
ionic solids and semiconductors and tend to give shorter lattices for alkali and alkali earth metals.
As with the S22 set, both the revPBE-vdW and rPW86-vdW2 yield equilibrium distances that are too long in most 
cases. This is improved by the functionals with optimized exchange: optPBE-vdW, optB88-vdW, and optB86b-vdW.
}
\label{sol_fun_comp}
\end{figure}

The lattice constants calculated with VASP are given in Table~\ref{tab_comp_sol} 
and shown as relative errors in Figure~\ref{sol_fun_comp}. 
For comparison we also give the errors of LDA, PBE, and  
the PBEsol functional, one of the GGA functionals~\cite{armiento2005,wu2006,madsen2007,perdew2008sol}
devised for solid-state calculations.
We also include the results of the adiabatic-connection fluctuation-dissipation theorem (ACFDT) 
in the random phase approximation (RPA)~\cite{langreth1975,gunnarsson1976,langreth1977} from Ref.~\onlinecite{harl2010}, 
which represents the state-of-the-art for solid state calculations. 
Before discussing 
the results in detail, let us just point out a striking 
feature of the results: the errors are not random and clear periodic trends are observed.  
All methods shown tend to give larger lattice constants for the transition 
metals, ionic solids, and semiconductors while the alkali and alkali earth lattices are too short.
This seems to correspond to the tendency of functionals to give larger lattice constants 
when going from left to right in the periodic table~\cite{haas2009}. 
This behavior does not seem to be improved by hybrid functionals~\cite{paier2006,schimka2011} and
is also present to some extent in the RPA lattice constants, although from the alkali and alkali
earth metals only the data for Na has been published~\cite{harl2010}.
In fact, even the functionals designed for solids do not lead to a qualitative improvement of the lattice constants. 
For example, the difference between the largest and the smallest relative errors
is similar for LDA, AM05, PBEsol, and PBE~\cite{haas2009}.

\begin{figure}[h!]
\centerline{
\includegraphics[height=8cm,angle=-90]{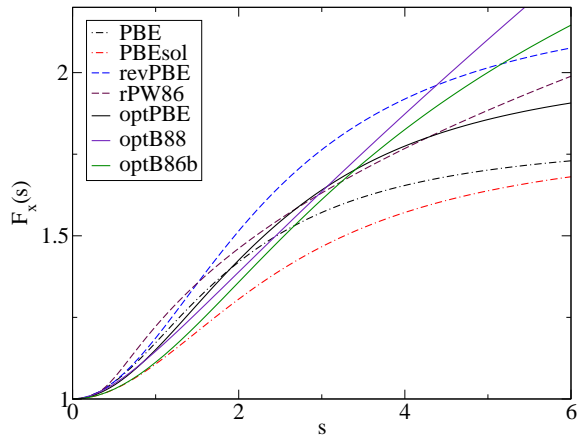}
}
\caption[Enhancement factors]
{The exchange enhancement factors $F_x$ of the functionals employed in this study: PBE, PBEsol, and revPBE which share
the same functional form but differ in the values of parameters, rPW86 which is used in the rPW86-vdW2~\cite{lee2010} functional, 
and three exchange functionals (optPBE, optB88, and optB86b) optimised for use with the vdW correlation~\cite{klimes2010}. 
The steepness for small reduced density gradients ($s$) is of crucial importance in determining the lattice constants. 
}
\label{sol_enhanc}
\end{figure}

Let us now discuss the results of the vdW functionals.  
The two van der Waals functionals proposed by the Langreth and Lundqvist groups (revPBE-vdW and rPW86-vdW2)
tend to give larger lattice constants (ME~$=0.106$~\AA\ for revPBE-vdW and ME~$=0.089$~\AA\ for rPW86-vdW2).
While revPBE-vdW overestimates all values, rPW86-vdW2, rather surprisingly, underestimates
the lattice constants of the alkali and alkali earth metals.
The errors are as large as 5.0\% (6.8\%) for revPBE-vdW (rPW86-vdW2) in the case of Ag and large for other
transition metals included as well as for Ge and GaAs.
The large errors are similar to the overestimation of the binding distance that has been observed before
for many systems~\cite{dion2004,puzder2006,chakarova2006,ziambaras2007}. 
This has been related to the too steep behavior of the exchange enhancement factor 
for small reduced density gradients which can be seen in Figure~\ref{sol_enhanc}. 
Although originally both revPBE and rPW86 exchange functionals were selected because they give similar binding to 
Hartree-Fock for some gas phase dimers, at short separations these functionals are too repulsive~\cite{murray2009}, 
which is important in hydrogen bonding and here for lattice constants.

The repulsion is largely decreased by utilising an exchange functional that has a less steeply rising $F_x$
and thus is less repulsive for short interatomic separations, 
such as the exchange functionals proposed in Ref.~\onlinecite{klimes2010} (see Figure~\ref{sol_enhanc}).
The optPBE-vdW is based on the PBE functional and it gives similar lattices to PBE for all the systems 
except for the alkali and alkali earth metals.
For these metals, the vdW correlation term gives better agreement with the reference than the semi-local
PBE correlation. 
The average errors are further reduced by using the optB88-vdW or optB86b-vdW functionals. 
However, for the
alkalis the lattice constants become too short and this worsens progressively as the ion size increases. 
This might be caused by overestimation of the dispersion energy in the vdW functional~\cite{vydrov2010pra}
or by the lack of higher order terms~\cite{lilienfeld2010}.
The optB88-vdW functional yields a mean error of $0.013$~\AA. 
The mean absolute error of $0.067$~\AA\ is comparable to the error of optPBE-vdW (MAE~$=0.064$~\AA). 
The optB86b-vdW which, like PBEsol, follows the limit of slowly varying density 
for small $s$, further improves the agreement with the reference (MAE~$=0.050$~\AA) and performs in between
PBEsol (MAE~$=0.033$~\AA) and PBE (MAE~$=0.067$~\AA).

Although it might be surprising at first sight that functionals optimized on interaction energies
of gas phase dimers give very good lattice constants, it just highlights the connection between 
the influence of the exchange part on lattice constants, molecular bonds, and intermolecular binding 
curves~\cite{tran2007, mattsson2009}.
In all these cases the small $s$ behavior is able to alter the properties, and by following the slowly varying electron gas limit
all these three measures tend to be improved.  

\begin{table*}[H]
\caption[Lattice constants of solids]
{Lattice constants in \AA\ of different solids calculated using VASP for different vdW functionals
and two GGA functionals (PBE and PBEsol). 
In addition, we show the LDA values taken from Ref.~\onlinecite{csonka2009}.
The values are compared to the experimental values corrected for zero-point energy effects (indicated by ``ZPEC")
also taken from Ref.~\onlinecite{csonka2009}. While both optPBE-vdW and optB88-vdW give mean absolute errors similar to 
those of PBE, this value for the optB86b-vdW functional is between those of PBE and PBEsol.
The original revPBE-vdW gives lattice constants that are too large.}
\label{tab_comp_sol}
\centerline{
\begin{ruledtabular}
\begin{tabular}{cccccccccc}
Exchange & revPBE & rPW86 &optPBE & optB88 &  optB86b& LDA &PBEsol &  PBE&  \\
Correlation& vdW&   vdW2 & vdW    &  vdW   &  vdW    & LDA &PBEsol & PBE&Exp. (ZPEC)\\
\hline
Cu&3.708& 3.757& 3.655&3.632&3.605&3.517 & 3.569 &3.635&3.595\\
Ag&4.254& 4.331&4.174&4.141&4.101&4.010  &4.059 &4.154&4.056\\
Pd&4.014& 4.086&3.960&3.941&3.909& 3.836 &3.876 &3.943&3.875\\
Rh&3.882& 3.945&3.843&3.831&3.805&3.755  &3.780 &3.830&3.793\\
\hline 
Li&3.453& 3.396&3.440&3.432&3.452& 3.363 & 3.436 &3.437&3.449\\
Na&4.233& 4.156 &4.195&4.169&4.191&4.054 & 4.174 &4.200&4.210\\
K&5.293&  5.177&5.225&5.168&5.202& 5.046 &5.216 &5.284&5.212\\
Rb&5.672& 5.550&5.584&5.506&5.541& 5.373 &5.572 &5.671&5.576\\
Cs&6.141& 5.987&6.022&5.899&5.945& 5.751 &6.015 &6.160&6.039\\
Ca&5.555& 5.493&5.502&5.450&5.465& 5.328 &5.461 &5.533&5.553\\
Sr&6.052& 6.005&5.979&5.917&5.921& 5.782 &5.913 &6.019&6.045\\
Ba&5.073& 5.058&4.987&4.917&4.906& 4.747 &4.894 &5.028&4.995\\
Al&4.084& 4.084&4.058&4.054&4.036& 3.985 &4.018 &4.041&4.020\\
\hline
LiF &4.116&4.080&4.067&4.033&4.037& 3.913 & 4.010 & 4.068&3.964\\
LiCl&5.223& 5.204& 5.153&5.114&5.103&4.968 & 5.067&5.152&5.056\\
NaF&4.752& 4.693 &4.693&4.647&4.658&4.502  &4.636&4.708&4.579\\
NaCl&5.750& 5.694& 5.673&5.622&5.627&5.465  &5.609&5.701&5.565\\
MgO&4.281& 4.282 &4.252&4.231&4.230&4.168  &4.222&4.257&4.184\\
\hline
C&3.600&3.608&3.585&3.577&3.572& 3.532  &3.557&3.574&3.543\\
SiC&4.406& 4.424& 4.386&4.375&4.369& 4.329 &4.354&4.377&4.342\\
Si&5.507& 5.523 &5.476&5.460&5.447&5.403  &5.429&5.465&5.416\\
Ge&5.864& 5.934 &5.793&5.762&5.725& 5.623 &5.680&5.766&5.640\\
GaAs&5.851&5.908 & 5.783&5.751&5.717&5.605  &5.667&5.752&5.638\\
\hline
ME (\AA)& 0.105& 0.088 & 0.050& 0.013& 0.010& $-0.100$ &$-0.005$&0.062 &\\
MAE (\AA)&0.105& 0.116 & 0.064 & 0.067&0.050 & 0.100 &0.033& 0.067&\\
MRE  (\%)& 2.3& 2.0  &1.1& 0.4 & 0.3& $-2.0$ & $-0.1$& 1.3&\\
MARE (\%)& 2.3& 2.6  &1.4& 1.4& 1.0& 2.0  &0.7 &1.4 &\\ 
\end{tabular}
\end{ruledtabular}}
\end{table*}

\section{Bulk moduli}
\label{secIV}

\begin{figure}[h!]
\centerline{
\includegraphics[height=9cm,angle=-90]{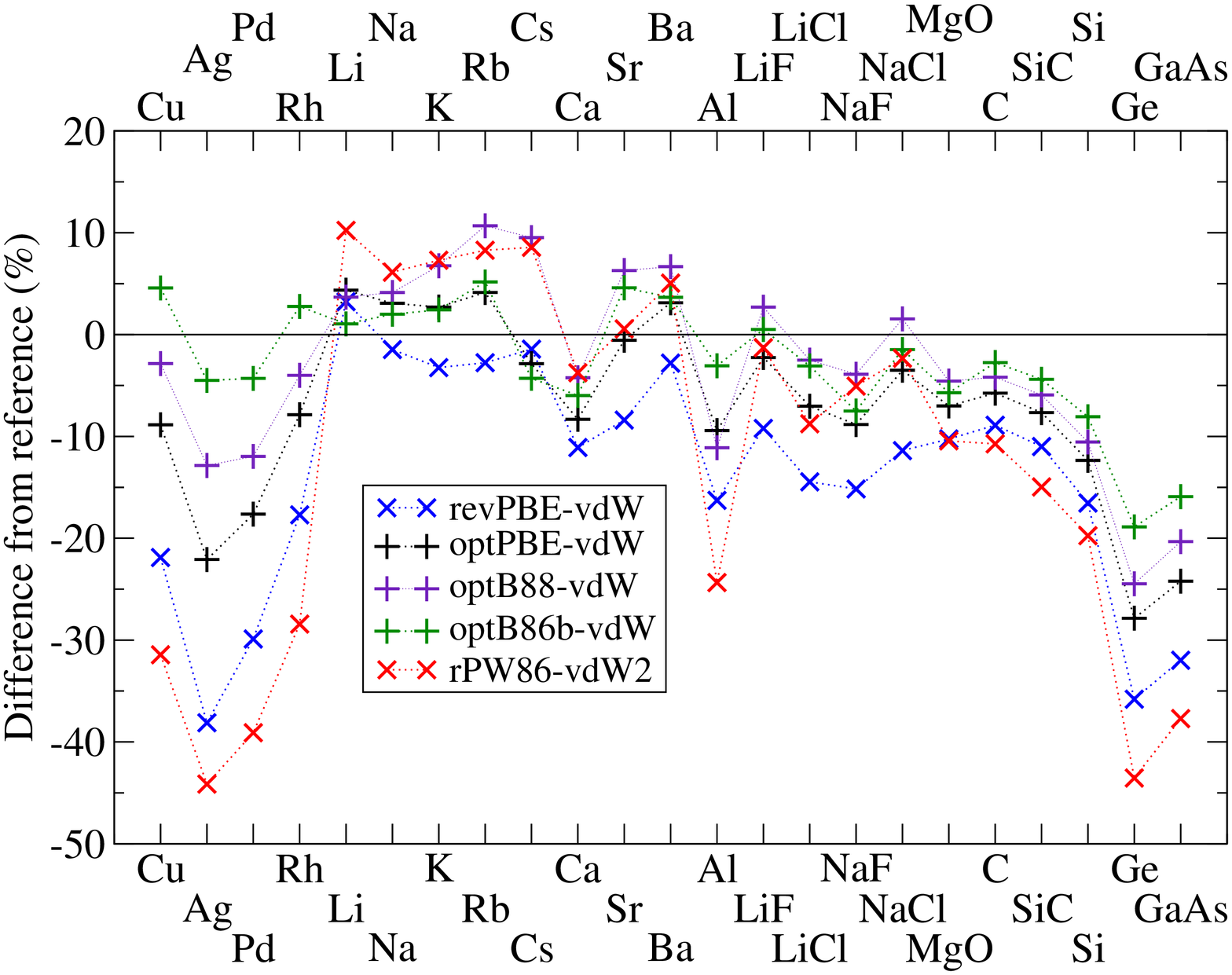}
\includegraphics[height=9cm,angle=-90]{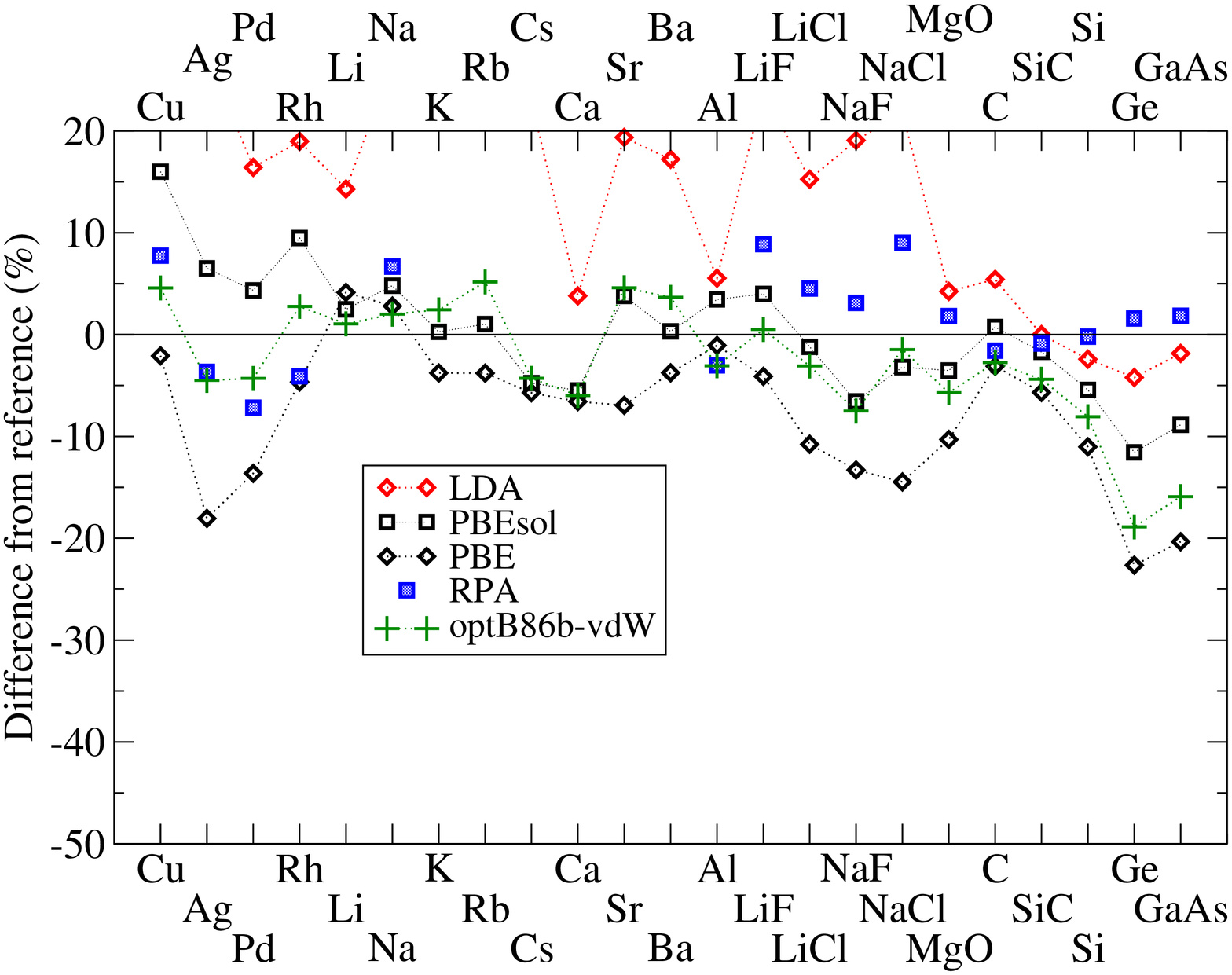}
}
\caption[Relative errors in bulk moduli]
{Relative errors in the bulk moduli for the methods considered in this study, results of the vdW functionals
are shown on the left, results of the other methods on the right where also the optB86b-vdW results were added.
The original revPBE-vdW and rPW86-vdW2 methods underestimate the moduli by up to 45\% and 
the optimized vdW functionals reduce the errors.}
\label{sol_fun_mod}
\end{figure}

It is known that the results of a given functional for bulk moduli are related to
the behavior for lattice constants. The shorter the predicted lattice constant, the higher the bulk modulus.
The vdW functionals tend to follow this trend as can be seen from the data in Table~\ref{tab_fun_mod}
and the relative errors shown in Figure~\ref{sol_fun_mod}. The revPBE-vdW and rPW86-vdW2 functionals 
give too soft lattices, with the bulk moduli smaller by more than $30$\% for Ag, Pd, Ge, and GaAs. 
This correlates well with the overestimation of the lattice constant by more than 3\% for these materials
with revPBE-vdW and rPW86-vdW2.  

There are several trends that one can observe, perhaps the clearest is the tendency of PBE and PBEsol
to underestimate the bulk modulus with the increase of the ion size. This is most prominent for semiconductors,
where it is clear that none of the vdW functionals alter this trend. On the other hand the RPA results
do not suffer this deficiency. Importantly, this softening trend for alkali metals is improved by the vdW functionals.
The reference experimental values were not adjusted for zero point energy effects which would slightly 
increase the reference values (up to $\sim$3\% in the case of Li~\cite{csonka2009}).
Let us then conclude that here again the optimized vdW functionals improve upon the original methods and they
follow the trend expected from the errors in the lattice constants. Specifically,
the average absolute errors increase in order PBEsol $<$ optB86b-vdW $<$ optB88-vdW $\approx$ PBE $\approx$ optPBE-vdW $<$
revPBE-vdW $<$ rPW86-vdW.

\begin{table*}[h]
\caption[Bulk moduli of the selected solids]
{Bulk moduli in GPa of the selected solids using different exchange correlation functionals 
with the LDA values taken from Ref.~\onlinecite{csonka2009}.
The experimental data are shown as well, these are not, however, corrected for
zero point energy effects which would lead to a slight increase of the values 
(see Ref.~\onlinecite{csonka2009}).}
\label{tab_fun_mod}
\centerline{
\begin{ruledtabular}
\begin{tabular}{cccccccccc}
Exchange & revPBE &rPW86 & optPBE & optB88 &  optB86b& LDA &PBEsol &  PBE&  \\
Correlation& vdW&vdW2 & vdW &vdW &vdW&LDA &PBEsol & PBE&Exp.\\
\hline
Cu&111&97 &129&138&149&190  & 165&139&142\\
Ag&67& 61 &85&95&104& 139   &116&89&109\\
Pd&137& 119&161&172&187&227  & 203&168&195\\
Rh&221& 193&248&258&276&320  & 295&256&269\\
\hline
Li&13.7& 14.7 & 13.9&13.8&13.4&15.2   &13.6&13.9&13.3\\
Na&7.39& 7.96 & 7.73&7.81&7.65& 9.50  &7.86&7.71&7.5\\
K&3.58&  3.97 & 3.80&3.95&3.79& 4.60  &3.71&3.56&3.7\\
Rb&2.82& 3.14 & 3.02&3.21&3.05& 3.54  &2.93&2.79&2.9\\
Cs&2.07& 2.28 & 2.04&2.30&2.01& 2.58  &2.00&1.98&2.1\\
Ca&16.4& 17.7 & 16.9&17.6&17.3& 19.1  &17.4&17.2&18.4\\
Sr&11.4& 12.5 & 12.3&13.2&13.0&  14.8 &12.9&11.5&12.4\\
Ba&9.04& 9.77 & 9.59&9.92&9.64& 10.9  &9.33&8.95&9.3\\
Al&67.1& 60.6 & 72.4&71.1&77.4& 83.8  &82.1&78.6&79.4\\
\hline
LiF&63.4&68.9&68.2&71.7&70.2&   86.5  & 72.6&66.9&69.8\\
LiCl&30.3& 32.3& 32.9&34.5&34.3& 40.8  &35.0&31.6&35.4\\
NaF&43.6&  48.8&46.9&49.4&47.5&  61.2  &48.0&44.6&51.4\\
NaCl&23.6& 26.0& 25.7&27.0&26.2& 32.4   &25.8&22.8&26.6\\
MgO&148&  148 &153&157&156&  172   &159&148&165\\
\hline
C&404&395&418&424&  431&467 &446&429&443\\
SiC&200& 191  &208&212&215& 225   &221&212&225\\
Si&82.8& 79.6  &86.9&88.7&91.2& 96.8  &93.8&88.3&99.2\\
Ge&48.7& 42.8  &54.7&57.3&61.5& 72.6  &67.0&58.6&75.8\\
GaAs&51.4& 47.1 & 57.3&60.2&63.6& 74.2  &68.9&60.2&75.6\\
\hline
ME (GPa)& $-15.9$&$-19.2$    &$-9.2$&$-7.0$ &$-4.2$ &10.3  &1.6 & $-7.3$& \\ 
MAE (GPa)& 15.9&  20.0  &  9.4& 7.6&5.6 &  10.9  & 5.1& 7.4& \\
MRE  (\%)& $-13.7$&  $-11.9$ & $-7.0$&$-3.4$ &$-3.0$&14.5   &0.1 & $-7.6$& \\
MARE (\%)& 14.0&  16.2  &  8.7& 8.2& 5.3&  15.3  & 5.2& 8.3& \\
\end{tabular}
\end{ruledtabular}}
\end{table*}

\section{Atomization energies of solids}
\label{secV}

\begin{figure}[h!]
\centerline{
\includegraphics[height=9cm,angle=-90]{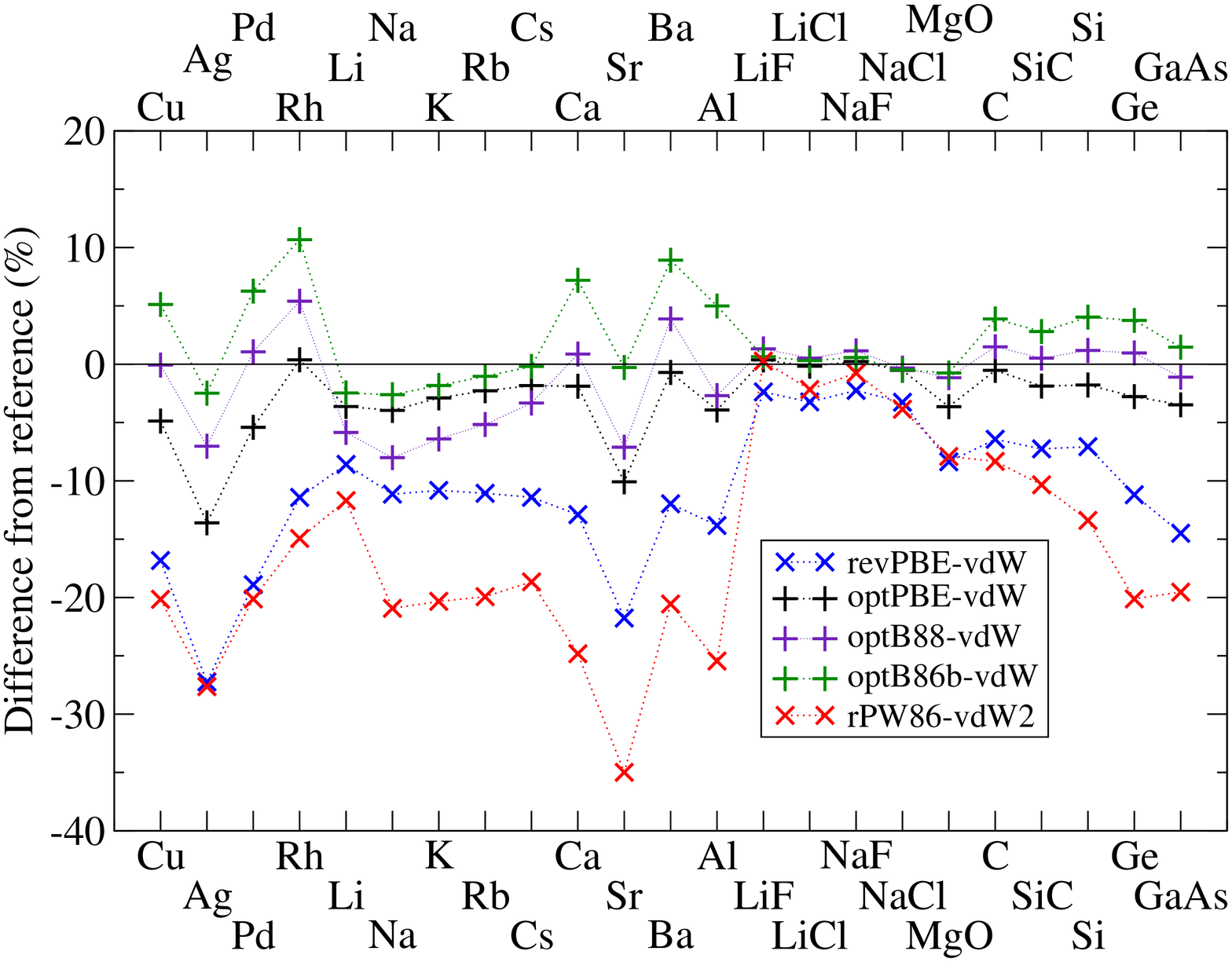}
\includegraphics[height=9cm,angle=-90]{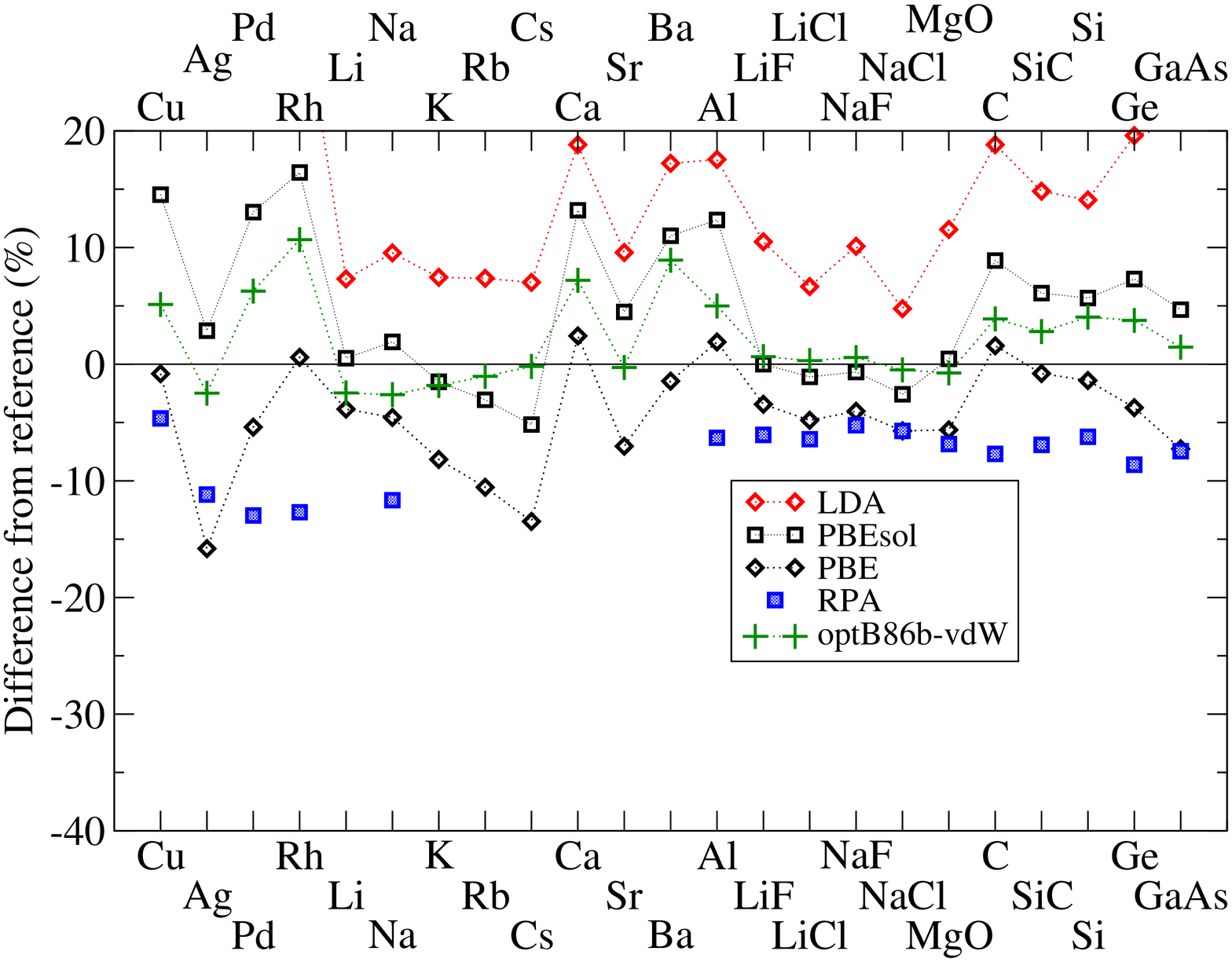}
}
\caption{Relative errors in atomization energies calculated using different DFT approaches. 
Data for various flavors of vdW functionals are shown in the left panel.
Data for LDA, semi-local PBE and PBEsol, the RPA method, and the optB86b-vdW functional are shown in the right panel.
The ZPE was subtracted from the experimental data. The optB88-vdW and optB86b-vdW tend to give values between those
of PBE and PBEsol for transition metals and semiconductors, in agreement with the behavior of their exchange enhancement factor.
However, they agree better with the reference for alkali halides, where even PBEsol underbinds. Moreover, they give consistent
errors for alkali metals where both PBE and PBEsol increasingly underbind with the increasing size of the ion.}
\label{atom_fun_comp}
\end{figure}

The calculated atomization energies for our selection of solids
are presented in Table~\ref{tab_comp_atom}, 
and the relative errors are shown in Figure~\ref{atom_fun_comp}. Again we include for comparison, LDA, PBE, 
PBEsol, and RPA data 
in Figure~\ref{atom_fun_comp}.
As can be seen the revPBE-vdW and rPW86-vdW2 functionals underestimate the atomization energies with average
relative errors of $-$11.1\% and $-$16.0\%, respectively. This underestimation is similar to the overestimation
of LDA (ME~$=15.1\%$). The optimized optPBE-vdW, optB88-vdW, and optB86b-vdW functionals 
give much improved results with the average relative errors of $-$3.0\%, $-$1.3\%, and 2.1\%, respectively. 
In most cases the optimized functionals tend to give larger atomization energies and increase in the order optPBE-vdW, optB88-vdW, 
and optB86b-vdW. 
Only in the case of the alkali metals does optB88-vdW give less binding than optPBE-vdW which 
for these materials agrees well with the reference values.

Interestingly, when one compares the GGA and vdW correlation functionals, there seems to be some systematic 
improvement as well, most notably for the alkali metals. While the PBE
atomization energies get progressively worse with the increase of the ion size, all the vdW functionals give
errors of a similar magnitude. While one can observe a similar trend for PBE atomization energies of semiconductors, 
which is decreased by the optimized vdW functionals, PBEsol seems to improve over PBE as well.
The atomization energies of the alkali halides calculated using the optimized vdW functionals are also in better 
agreement with the reference data than either PBE or PBEsol. 
The effect of different correlation functionals will be discussed more in the next section.

\begin{table*}[h]
\caption[Atomization energies of solids calculated using vdW functionals]
{Atomization energies in eV for various solids calculated using VASP for different exchange-correlation functionals.
We show the data of the revPBE-vdW and rPW86-vdW2 functionals,
the optimized vdW functionals, and results of LDA, PBEsol, and PBE.
The values are compared to the experimental values corrected for zero-point energy effects
taken from reference~\cite{csonka2009}. All three optimized functionals give better results than
either PBE or PBEsol. The LDA values were taken from 
Ref.~\onlinecite{harl2010} for semiconductors, ionic solids, transition metals, and Al. The atomization
energies of alkali and alkali earth metals were taken from Ref.~\onlinecite{perdew2009rev}.}
\label{tab_comp_atom}
\centerline{
\begin{ruledtabular}
\begin{tabular}{cccccccccc}
Exchange & revPBE & rPW86 &optPBE &  optB88& optB86b & LDA&PBEsol & PBE & \\
Correlation& vdW&vdW2 & vdW & vdW & vdW & LDA &PBEsol & PBE &Exp. (ZPEC)\\
\hline
Cu&2.93&  2.81&3.35&3.52&3.70& 4.55 &4.04&3.49&3.52\\
Ag&2.16&  2.15&2.57&2.76&2.90& 3.64 &3.06&2.50&2.97\\
Pd&3.18&  3.13&3.71&3.96&4.16& 5.08 &4.43&3.71&3.92\\
Rh&5.12&  4.92&5.81&6.10&6.40& 7.67 &6.73&5.82&5.78\\
\hline
Li&1.52& 1.47 &1.61&1.57&1.63& 1.79 &1.68&1.60&1.67\\
Na&1.01& 0.90 &1.09&1.04&1.10& 1.24 &1.15&1.08&1.13\\
K&0.84&  0.75 &0.91&0.88&0.92& 1.01 &0.93&0.86&0.94\\
Rb&0.76& 0.69 &0.84&0.81&0.85& 0.92 &0.83&0.77&0.86\\
Cs&0.72& 0.66 &0.80&0.79&0.81& 0.87 &0.77&0.70&0.81\\
Ca&1.62& 1.40 &1.82&1.88&1.99& 2.21 &2.11&1.90&1.86\\
Sr&1.36& 1.13 &1.56&1.61&1.73& 1.90 &1.81&1.61&1.73\\
Ba&1.68& 1.52 &1.90&1.99&2.08& 2.24 &2.12&1.88&1.91\\
Al&2.96& 2.56 &3.30&3.34&3.61& 4.04 &3.86&3.50&3.44\\
\hline
LiF&4.36& 4.48 &4.49&4.53&4.50& 4.94 &4.47&4.32&4.47\\
LiCl&3.47&3.51 & 3.58&3.61&3.60&3.83  &3.55&3.42&3.59\\
NaF&3.89& 3.95 &4.00&4.02&4.00& 4.38 &3.95&3.82&3.98\\
NaCl&3.23&3.21 & 3.32&3.33&3.32& 3.50 &3.25&3.15&3.34\\
MgO&4.83& 4.85 &5.08&5.21&5.23& 5.88 &5.29&4.97&5.27\\
\hline
C&7.09& 6.95 &7.54&7.70&7.88& 9.01  &8.26&7.70&7.58\\
SiC&6.02& 5.82& 6.37&6.52&6.67& 7.45  &6.88&6.44&6.49\\
Si&4.35&  4.05&4.60&4.74&4.87& 5.34  &4.95&4.62&4.68\\
Ge&3.43&  3.09&3.76&3.90&4.01& 4.62  &4.15&3.72&3.86\\
GaAs&2.90& 2.73 &3.27&3.36&3.44&4.09   &3.55&3.15&3.39\\
\hline
ME (eV)& $-0.34$&   $-0.46$ & $-0.09$& 0.00  & 0.10&  0.56 &0.20&$-0.11$ &\\
MAE (eV)&0.25 &   0.46 & 0.09 & 0.07 &0.12 &   0.56 &0.22& 0.13&\\
MRE  (\%)& $-11.1$&  $-16.0$&  $-3.0$& $-1.3$ & 2.1&  15.1  &4.7& $-4.4$&\\
MARE (\%)& 11.1&   16.0& 3.1&  2.9& 3.2&   15.1  &6.0 &5.0 &\\
\end{tabular}
\end{ruledtabular}}
\end{table*}

\section{The effect of non-local correlation}
\label{secVI}

\begin{figure}[h!]
\centerline{
\includegraphics[height=9cm,angle=-90]{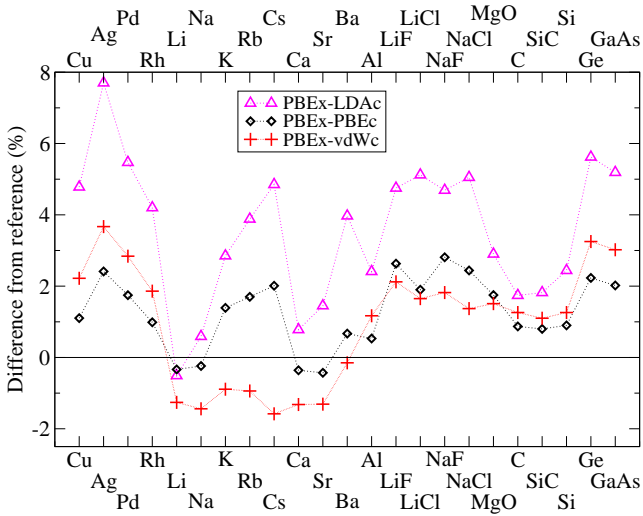}}
\caption{Comparison of the relative errors in the lattice constants for the PBE exchange functional
with LDA, PBE, and vdW correlation functionals. 
}
\label{sol_pbe_vdw}
\end{figure}

\begin{figure}[h!]
\centerline{
\includegraphics[height=9cm,angle=-90]{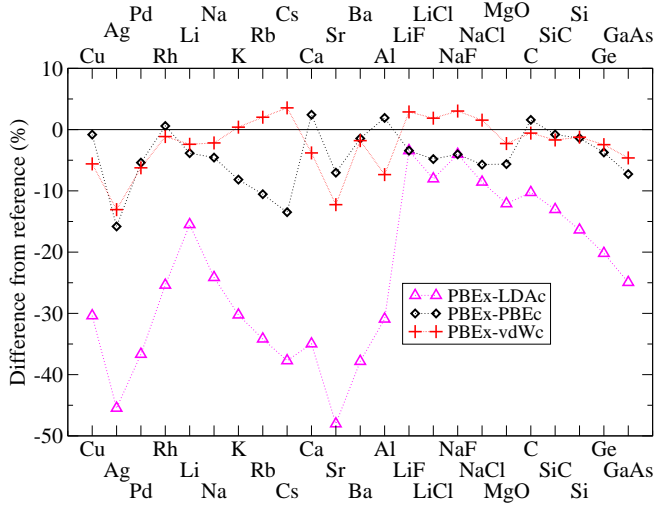}}
\caption{Relative errors in the atomization energies for the PBE exchange functional combined with LDA, PBE, and vdW
correlation functionals.}
\label{atom_pbe_vdw}
\end{figure}

Although we know that the vdW-DF correlation form is only approximate, it is interesting
to see what changes occur when semi-local correlation such as the PBE correlation
(referred to as ``PBEc") is replaced by the non-local form of vdW. 
(The ``vdW correlation" is the $E_{\rm c}^{\rm LDA}+E_{\rm c}^{\rm nl}$ correlation energy). 
To study this change we have calculated the lattice constants using the PBE
exchange functional (referred to as ``PBEx") and LDA, PBE, and vdW correlation functionals. 
This way we can directly compare the effect of adding PBE semi-local or vdW non-local corrections. 
Let us first present the results for the lattice constants in Figure~\ref{sol_pbe_vdw}. 
At first sight, the PBEx-PBEc and PBEx-vdWc give rather similar results, consistently 
decreasing the PBEx-LDAc lattice constant\cite{foot_LDA}.
This means that an exchange functional which gives good results for solids with PBE correlation
will tend to give good results with the vdW correlation as well.
As we noted before, there is, however, a clear difference for the alkali metals, 
where the PBEx-PBEc gives progressively worse lattice constants with the increase of the ion size.
This is even more pronounced for the atomization energies, shown in Figure~\ref{atom_pbe_vdw}, 
where PBEx-PBEc underbinding starts at only $-3$\% for Li but worsens to $\sim$$-15$\% for Cs.
This trend is clearly reversed by the vdW correlation, although too much.
The tendency to underbind larger ions by semi-local functionals is analogous to the behavior of semi-local
functionals for noble-gas dimers in the gas phase~\cite{mourik2002}. 
The interaction energy is obtained only from the region of electron density overlap and therefore does not
scale in the same way as the size of the ion.

Careful observation reveals that the differences between PBE and vdW correlations are 
qualitatively similar for transition metals and semiconductors. The lattice constants
are larger by $\sim$1\% (except for the smaller semiconductors, where the difference
is less) while the atomization energies have a smaller range of errors.
It has been shown~\cite{harl2009,harl2010} using RPA that out of the Rh, Pd, Ag group, 
electron correlation beyond Hartree-Fock-like exchange is the most important for Ag.
With this in mind, it is satisfactory to see the decreased range of errors in atomization energies for this group, 
although the trend is not cured completely. 
Note that the trend in lattice constants is not improved and 
it would be actually worsened by using a hybrid functional~\cite{paier2006}.
Thus the semi-local PBE or non-local vdW correlation with semi-local or hybrid exchange seems to be 
unable to describe the delicate balance of the interactions in the late transition metals.
For semiconductors PBEx-vdWc further increases the lattice constants compared to PBEx-PBEc.
Since for these systems the non-local correlation is less important than the semi-local contribution, 
the results suggest that the semi-local part of the vdW correlation is effectively less attractive than PBEc.
However, PBEx-PBEc worsens 
the atomization energies for solids with larger atoms where the vdW correlation improves the trend.

\section{Discussion and Conclusions}
\label{secVII}

In this study we have compared solid state properties obtained with different semi-local and non-local 
exchange-correlation functionals and we summarize the main results in Table~\ref{tab_sol_final}. 
We have found that the particular choices of exchange functionals made by Langreth and Lundqvist and co-workers
for the revPBE-vdW and rPW86-vdW2 functionals (\ie, revPBE and rPW86)
lead to large overestimations of lattice constants and underestimations of bulk moduli
and atomization energies for most of the solids considered.
In addition, the errors have a wide range, \eg, rPW86-vdW2 
underestimates the lattice constant of Li by 1.5\% but overestimates the value for Ag by 6.7\%.
The atomization energies are underestimated by more than 0.3~eV on average.
The optimized exchange functionals introduced in Ref.~\onlinecite{klimes2010} (\ie, optPBE-vdW and optB88-vdW)
and the optB86b-vdW functional introduced here improve over revPBE-vdW and give lattice constants
that are similar to those of PBE. 
This leads to similar improvements for the bulk moduli.
From our study it seems that the vdW correlation functional does not improve dramatically over PBE 
except for the lattice constants of the alkali metals.
This means that there is still some spread of the errors in the lattice constants which is not improved 
compared to PBE or PBEsol and further developments are required to reduce this range of errors.
Importantly, the atomization energies seem to be qualitatively improved when a non-local correlation functional is used.
This is most notable for the alkali metals, where PBE and PBEsol increasingly underbind 
with the increasing size of the ion but the vdW functionals suffer no such deficiency. 
Moreover, the atomization energies of ionic solids are in very good agreement with the 
experimental values.

\begingroup
\squeezetable
\begin{table}[h]
\caption[Results]
{Summary of the results for lattice constants and atomization energies obtained using PBE, 
the revPBE-vdW and rPW86-vdW2 functionals of Langreth and Lundqvist and co-workers, 
and using the vdW functionals with optimised exchange.}
\label{tab_sol_final}
\centerline{
\begin{tabular}{lccc}
\hline
\hline
                  & PBE & revPBE-vdW, rPW86-vdW2&  Optimized exchange\\
\hline
Lattice constants      & &Worse than PBE & Similar to or better than PBE\\
MARE              &1.4\% &  2.3\% (revPBE-vdW)     & 1.4\% (optB88-vdW)    \\
\hline
Atomization energies  & &Worse than PBE & Better than PBE \\
MARE               &5.0\%    &     11.1\% (revPBE-vdW)  &  2.9\% (optB88-vdW)   \\
\hline
\hline
\end{tabular}
}
\end{table}
\endgroup

Let us now discuss the results obtained here in a broader context. First, after the local and semi-local
approximations, the non-local density functionals are the next logical step before the orbitals are 
introduced in the exchange-correlation energy such as is done in the RPA or hybrids.  
In this sense the non-local correlation functionals offer great promise. 
However, both revPBE-vdW and rPW86-vdW2 suffer from too 
much repulsion at short distances, a well known feature of revPBE-vdW 
for systems like the gas phase dimers~\cite{dion2004,puzder2006,lee2010}.
In this study we have shown that lattice constants of solids are subject to similar errors.
This is a significant problem since accurate lattice constants are crucial for the predictive
power of theory~\cite{feibelman2008}.
One possible way to alleviate the problems is to change the exchange functional.
We have shown that functionals with less steeply rising exchange enhancement factors
for small $s$ will improve both results on the S22 set and, in this study, the lattice
constants of solids. 
However, it is known that this change will reduce the accuracy for molecular atomisation energies
and further development to alleviate this is required.

At a more fundamental level, the question of what form of exchange and correlation to use is still to be resolved.
In principle one can try to find an exchange functional compatible with ``dispersionless"
interaction energies~\cite{pernal2009}. 
Another approach might be to fit a functional
to interaction energies based on the so called exact exchange (EXX) in the ACFDT formalism.
This would allow the correlation part to be compared directly to the ACFDT correlation
energy (\eg, in the RPA approximation). 
However, this might not be qualitatively that different from trying to reproduce HF binding curves. 
Moreover, the EXX energy depends on the single particle orbitals
and the correlation part will be just $E_{\rm c}^{\rm EXX}=E({\rm exact})-E({\rm EXX})$. 
Therefore no ``exact" correlation energy can be defined in this sense.
Even defining $E_{\rm c}^{\rm EXX}$ using some choice of orbitals will mean that this needs to be
reproduced by a given DFT functional which seems to be rather difficult. 
So far, even the form of semi-local correlation that should be used is an 
ongoing debate~\cite{dion2004,vydrov2008}.
This includes the question of how much of the semi-local correlation energy the vdW correlation functional recovers.
In this light, there is a need for reference systems to help the development, similar to GGA functionals
where lattice constants, atomization energies, bond lengths, and other data have been extremely
useful.
The doubts and discussions concerning the vdW functionals just
highlight the need for accurate reference data for gas phase clusters, adsorbates, solids, and so on.
The approach of using quantum chemistry methods for the solid 
state~\cite{paulus1999,li2008,nolan2009,tosoni2010,jenness2010,grueneis2010}
is one that seems very useful and deserves more attention.

To conclude, we have calculated solid state properties of a set of solids using a self-consistent
implementation of the vdW-DF method in the VASP code. We have shown that the method agrees well
with all-electron data which are much more time consuming to obtain. 
The lattice constants of solids are, in analogy to what has been reported for gas phase dimers~\cite{dion2004,puzder2006,lee2010}, 
too large with the original revPBE-vdW and rPW86-vdW2 methods but 
improved when optimized exchange functionals are used.
Indeed, optB86b-vdW gives errors in lattice constants between PBE and PBEsol and thus yields 
accurate binding properties for gas phase clusters and also describes bulk materials well.
The atomization energies of solids are considerably improved when the optimized functionals are used.
This work provides some clear reference data as to how the vdW-DF family of functionals perform which
should be useful in the further development of the method.

\begin{acknowledgments}

A. M. is supported by the EURYI award scheme (see: www.esf.org/euryi), the EPSRC, and the European Research Council.
DRB is supported by the Royal Society.
We are grateful to the London Centre for Nanotechnology and UCL Research Computing
for computational resources.
Via our membership of the UK's HPC Materials Chemistry Consortium, which is funded by EPSRC (EP/F067496), 
this work made use of the facilities of HECToR, the UK's national high-performance computing service, 
which is provided by UoE HPCx Ltd at the University of Edinburgh, Cray Inc and NAG Ltd, 
and funded by the Office of Science and Technology through EPSRC's High End Computing Programme.

\end{acknowledgments}

\appendix

\section{optB86b}
\label{app_optb86b}

Here we briefly present the optB86b exchange functional, a more in depth discussion will be published 
elsewhere. 
From detailed studies of the exchange functionals and binding curves it became apparent that the 
behavior of the exchange enhancement factor ($F_x$) for small reduced density gradients ($s$) affects the
position of the repulsive Pauli wall. Functionals with steeply increasing $F_x$ are more repulsive, 
and more importantly, start to be repulsive for longer distances than functionals with $F_x$ less
steep or flat, like LDA. 
Therefore lattice constants tend to be longer when one goes from LDA to PBEsol to PBE to 
revPBE~\cite{mattsson2008,haas2009} and similar observations can be made for equilibrium distances
of gas phase clustes, \eg\ the water dimer~\cite{santra2007,mattsson2009}.
This has been exploited in the PBEsol functional which decreases the average overestimation of the 
PBE equilibrium distances~\cite{perdew2008sol}.
In an analogous way the overestimation of the revPBE-vdW binding distances, observed for many systems, 
can be reduced by choosing a functional that rises less steeply for small $s$.
Using the same small $s$ behavior as PBEsol leads to a good agreement of the gas phase dimer
binding curves with the reference data. 
For large $s$, it has been suggested that $F_x$ should have $s^{2/5}$ behavior~\cite{murray2009}.
We modified the B86b exchange functional to obey these limits (although the second with a coefficient 
slightly different from the one suggested in Ref.~\onlinecite{murray2009}). The form of the optB86b functional
is then 
\begin{math}
F_x^{optB86b}=1+{\mu s^2\over(1+\mu s^2)^{4/5}}\,, \mu=0.1234\,,
\end{math}
and the function is shown in Figure~\ref{sol_enhanc}.
The optB86b-vdW gives almost the same results on the S22 dataset as the optB88-vdW functional, 
namely the mean absolute deviations are 12~meV for the total set and 13, 16, 6~meV for the hydrogen, 
dispersion, and mixed bonding subsets (using the reference data of Podeszwa~{\it et al.}~\cite{podeszwa2010} 
on the geometries of Jure\v{c}ka~{\it et al.}~{\cite{jurecka2006}}).
However, since this form has a less steeply rising $F_x$ for large $s$ than optB88, it is less
repulsive for distances larger than optimum. 
This leads to a smaller error cancellation between exchange and the overestimated correlation
than for the optB88-vdW functional.

\section{All electron density based lattice constants}

Our tests comparing the approximate vdW evaluation in VASP to all-electron calculations show
very good agreement between both approaches. However, it is not clear if reference quality 
calculations can be performed since the vdW energy depends on the PAW potential used.
In this part we aim to obtain all-electron based lattice constants and then assess the accuracy 
of VASP against this benchmark.
We start by showing that by utilising different PAW potentials the lattice constants 
differ. For example, we show lattice constants of Ge evaluated with three different PAW potentials in
Table~\ref{tab_comp_ge}. 
The potentials are Ge with 4 valence electrons, Ge\_d with 14 electrons, 
and a hard Ge\_h with 14 electrons. 
One can see that the differences between the optB86b-vdW lattice constants cannot be completely
attributed to the differences caused by the PAW potential, shown by the optB86b-LDA values.
Therefore in the following we first test convergence of the various parameters involved.
For this we use Ge because of its medium size and the fact that three different PAW potentials are
available for it.
Later, in Section~\ref{comp_app} we obtain the all-electron based data for the whole set.

\begin{table}[h]
\caption{Lattice constants in \AA\ of Ge evaluated using various approximations for the $E_c^{nl}$
for the optB86b-vdW functional.
Three PAW potentials were used, Ge has four valence electrons, Ge\_d and Ge\_h fourteen, and
Ge\_h has a smaller core radius. The FFT grid contains 120 points in each direction, so that the
grid spacing in the cell is $\sim$0.05~\AA. All the calculations used $\qcut=10$
and $N_\alpha=30$ to allow for a comparison between VASP and the all-electron results.
The differences in the optB86b-LDA lattice constants represent the error given by the PAW potential.
One can see that the all-electron based evaluations of the vdW energy ($\varrho^{cut 20}_{ae}$ and $\varrho_{ae}^{no soft}$)
give almost the same differences in lattice constants between the different PAW potentials as optB86b-LDA.
The optB86b-vdW lattice constant calculated with VASP agrees well with the all-electron calculations
for the hard potential, the agreement is worse for the Ge and Ge\_d potentials.
However, in the worst case of the Ge potential this deviation is 0.018~\AA, much smaller
than the difference of $\sim$0.07~\AA\ when only the real valence density ($\varrho_{val}$) is used.}
\label{tab_comp_ge}
\centerline{
\begin{ruledtabular}
\begin{tabular}{cccc}
     &Ge    & Ge\_d  & Ge\_h  \\
\hline
optB86b-LDA &5.857& 5.842 & 5.845\\
optB86b-vdW & 5.764& 5.714& 5.726 \\
$\varrho_{val}$& 5.814& 5.735& 5.738\\
$\varrho^{cut 20}_{ae}$ & 5.746 & 5.729&5.732  \\
$\varrho_{ae}^{no soft}$&5.740 & 5.723&5.726  \\
\end{tabular}
\end{ruledtabular}
}
\end{table}

To calculate the all-electron vdW energy we use a standalone programme based on the vdW routines in 
SIESTA~\cite{soler2002,artacho2008,soler2009}.
However, the calculation of the AE based lattice constants is not straightforward as several parameters
need to be converged. Importantly, the AE density represented on a finite grid leads to numerical errors close
to the ionic cores. Therefore we first smoothly cut the electron density around the cores and test the convergence 
of the other parameters. We then study the effect of the cut of the density. Furthermore, we test if the
lattice constant can be evaluated using only the valence electron density. We use the lattice constant of 
Ge with the Ge\_h PAW data set for the tests of the parameters.

\subsection{Convergence tests}

The efficient vdW algorithm introduces two basic parameters that control the quality of interpolation of the
$q_0$ function: a cut-off $\qcut$ and number of interpolation points $N_\alpha$. The vdW energy also 
depends on the underlying FFT grid and the density cut-off needed to avoid numerical errors close to the cores.
The FFT grid is the most straightforward parameter to converge and we find that grid spacing around 0.03~\AA\ can
be considered converged, compared to grid spacing of 0.04~\AA\ the lattice constant changes by only 0.002~\AA.

\begin{table}[h]
\caption{Dependence of the Ge lattice constant on the number $N_\alpha$ of $q_0$ interpolation points and the
$\qcut$.
The electron density was smoothed above 20~a.u. and a very fine FFT grid with 200 points in each direction
(corresponding to 0.03~\AA\ spacing).}
\label{tab_ge_qcna200}
\centerline{
\begin{ruledtabular}
\begin{tabular}{ccccc}
$\qcut$     & & $N_\alpha$& & \\
     &20    & 30  & 40& 50  \\
\hline
5    & 5.751& 5.749 &5.748&5.748\\
10   & 5.740& 5.734& 5.731&5.730 \\
18   & 7.505& 5.733& 5.732&5.732  \\
\end{tabular}
\end{ruledtabular}}
\end{table}

The values of $\qcut$ and $N_\alpha$ affect the value of the lattice constant more significantly. 
As can be seen in Table~\ref{tab_ge_qcna200}, 
the lattice constant seems to converge when both $\qcut$ and $N_\alpha$ are increased. 
The values of $N_\alpha=30$, $\qcut=10$ give results that are in a very good agreement with
the lattice constants obtained with either $N_\alpha$ or $\qcut$ increased. 
Moreover, the $\qcut=10$ values are almost identical to the values obtained with $\qcut=18$.
We therefore now set $\qcut=10$ and study how the lattice constant depends on the density cut-off and $N_\alpha$.
Table~\ref{tab_ge_dna} shows that the values first converge when $N_\alpha$ is increased up to 40, further
increase, to $N_\alpha=50$ and $N_\alpha=80$, gives oscillating values. 
This is more pronounced
with higher density cut-offs. This seems to be caused by ``overinterpolation" of the $q_0$ function, 
which would be probably less severe with an even denser grid. However, there seems to be no point
in doing this since the calculations using $N_\alpha=40$, $\qcut=10$ agree with more stringent 
settings to within 0.001~\AA.
Therefore, to obtain the reference lattice constants, 
we use density cut-off 20 a.u., $N_\alpha=40$, $\qcut=10$, and FFT grid with fine spacing around 0.03~\AA.
Since increasing $N_\alpha$ is computationally demanding, we use the values $N_\alpha=30$, $\qcut=10$ in VASP
calculations.

\begin{table}[h]
\caption{Lattice constant of Ge calculated on the very fine grid with 200 points in each direction
for different cut-offs of the all-electron density and different $N_{\alpha}$ interpolation points.
The cut-off  value for $q_0$ was set to $\qcut=10$. The change of the lattice constant is small
overall when the density cut-off is increased for a constant $N_{\alpha}$. The variation is higher
for constant cut-off and increasing number of interpolation points. For the highest density cut-off 
and the highest number of interpolation points the data deviates because of an insufficient real space grid.}
\label{tab_ge_dna}
\centerline{
\begin{ruledtabular}
\begin{tabular}{cccccc}
$\varrho_{\rm cut}$~(a.u.)     & & $N_\alpha$& &  &\\
     &20    & 30  & 40& 50  & 80 \\
\hline
20    & 5.740& 5.734&5.731&5.730& 5.733\\
100   & 5.734&5.734 &5.731 &5.725&5.742 \\
1000  & 5.738& 5.733&5.730 &5.728&5.747  \\
\end{tabular}
\end{ruledtabular}}
\end{table}

Before calculating the lattice constant on the whole solid state test, we present two 
alternative approaches to the calculation.
First, it turns out that the problematic part of the calculation that makes the direct evaluation
of the vdW energy impractical is the soft correction term, introduced in Ref.~\onlinecite{soler2009}. 
However, the lattice constants calculated with or without this term are
virtually identical when the electron density is cut and a ``hard" vdW kernel is used
(\cf\ the data in Table~\ref{tab_ge_dnanc} and Table~\ref{tab_ge_dna}). 
Therefore one can use the all-electron density to evaluate the lattice constant if the soft 
correction is not added. However, as the data for large $N_\alpha$ in Table~\ref{tab_ge_dnanc} suggest, 
there is some numerical noise introduced from the interpolation as well.
Despite this, the difference in the lattice constants of the all-electron smoothed density 
and the all-electron density without the soft correction is very small ($<0.1$\%).

\begin{table}[h]
\caption{Dependence of the Ge lattice constant of the all-electron density cut-off and the number
of interpolation points $N_{\alpha}$. The soft correction was not added to $E_c^{nl}$. $\qcut=10$ was used. 
In this case the all-electron density can be used without any cut-off (row $\infty$)
since the contribution from the inner shells is small. 
The agreement with data in Table~\ref{tab_ge_dna}
is almost perfect, with the exception of the lattices obtained with $\varrho_{\rm cut}=1000$.}
\label{tab_ge_dnanc}
\centerline{
\begin{ruledtabular}
\begin{tabular}{cccccc}
$\varrho_{\rm cut}$~(a.u.)  & & $N_\alpha$& & &\\
     &20    & 30  & 40& 50  &80\\
\hline
20    &5.736 & 5.734& 5.731& 5.730&5.733\\
100   &5.731 &5.734 &5.731 &5.725&5.742 \\
1000  &5.732 &5.731 &5.728 &5.726& 5.743\\
$\infty$ & 5.732  & 5.731& 5.728&5.726&5.743 \\
\end{tabular}
\end{ruledtabular}}
\end{table}

It is also interesting to try to use the real valence density (\ie\ not the pseudo valence density) to obtain the
lattice constant. In this case the lattice constant converges quickly with all the parameters and, in the case of Ge, 
the converged value is 5.737~\AA, slightly larger than the 5.732~\AA\ obtained with the all-electron density.
As we shall see, the agreement for other materials strongly depends on the number of electrons included in the valence shell.

Let us now summarize the results and compare the lattice constants obtained using the softer
potentials as well. 
We compare the VASP calculations to the real valence,
and all-electron with and without electron density cut-off in Table~\ref{tab_comp_ge}.
Since we want to be able to compare to our VASP implementation we use $N_{\alpha}=30$ and $\qcut=10$.
One can notice that the
changes in the all-electron lattice constants ($\varrho^{cut 20}_{ae}$ and $\varrho_{ae}^{no soft}$)
correspond well to the respective changes in the optB86b-LDA values. The real valence
calculation ($\varrho_{val}$) without the $d$ electrons gives too a large lattice constant, the difference
is almost halved in the case of the VASP calculation where the 
partial electronic core charge is added as well.
Therefore we can expect a very good agreement of the VASP and all-electron data for PAW potentials
that are either hard or contain more than one electronic shell.

\subsection{Comparison on the whole set}
\label{comp_app}

Now we proceed to calculate the lattice constants of the chosen solids using the approaches shown above.
This allows us to test if we need to use the quite cumbersome all-electron evaluation or if VASP
can be used. 
We use several ways to estimate the lattice parameter: first, we calculate it directly from VASP using
the approximated $E_c^{nl}$. Second, the real valence density is used, third, the all-electron density without
the soft correction. Finally, the vdW energy is calculated on the all-electron density with
a density cut-off imposed to avoid numerical errors. The VASP implementation uses $N_\alpha =30$ interpolation
points with $\qcut=10$, the real valence and all-electron calculations use $N_\alpha=40$
interpolation points and $\qcut=10$.
The number of FFT grid points is set by hand to a large number so that the integration grid spacing is $\sim$0.03~\AA.

\begin{figure}[h!]
\centerline{
\includegraphics[height=4.0cm]{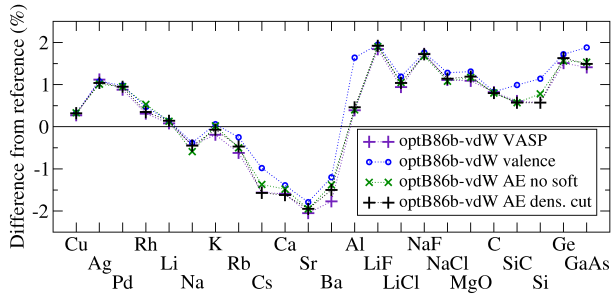}}
\caption[Comparison of VASP and all-electron lattice constants]
{Lattice constants of various solids calculated with different approximations of the
non local van der Waals energy for the optB86b-vdW functional. The self-consistent implementation
in VASP and non-self-consistent calculations based on the same density are reported.
These use the real valence density (``valence") and the all-electron density without (``AE no soft")
and with (``AE dens. cut") the soft correction. The VASP calculations tend to give better agreement with
the AE calculations because of the partial electronic core charge density added to the pseudo valence density.}
\label{sol_ae_comp}
\end{figure}

The lattice constants using optB86b-vdW are collected in Table~\ref{tab_comp_sol2}
and shown in Figure~\ref{sol_ae_comp}. One can see that the data obtained with VASP (violet $+$) 
are in a very good agreement with the all-electron calculations (black $+$) that use smoothed electron density.
This then justifies the approximations involved in the evaluation of the vdW correlation energy in VASP.
The calculations without the soft-correction (green $\times$) on the all-electron density are generally very
similar to the smoothed density calculations. As we have shown earlier, this seems to come more from the
representation of the electron density on the finite grid.
We find the largest deviations from the AE results for alkali and alkali earth metals,
which are very sensitive to the errors in the vdW correction because of their small bulk moduli.
This means that the lattice constant calculated with the all-electron density has some error and the very good agreement
between the VASP and AE calculations for Cs might be accidental, the trend 
towards shortening the lattices with the increased size of the ion is not affected.
For the other materials the differences are below 0.1\%, \ie, 
the results differ only at the third decimal place in most cases. This level of accuracy of our VASP calculations
is more than sufficient to recover the trends and also similar to or better than the differences for the same solid and
functional obtained with different codes~\cite{csonka2009}.

An interesting approach which would circumvent the problematic calculation of the AE vdW correction would be to use
the real valence density to calculate $E_c^{nl}$.
However, lattice constants obtained with this approach (blue circles in Figure~\ref{sol_ae_comp})
are slightly larger than the AE ones.
This difference seems to crucially depend on the number of shells included in the valence, this is supported by the
fact that the largest errors are observed for Al (3 electrons in valence), Si (4), and As (5).
Although the valence electron density based data can't be used to obtain a reliable lattice constant in some cases, they
seem to give a good upper bound of the AE based values.

\begin{table}[h]
\caption{Lattice constants in \AA\ calculated with the optB86b-vdW functional using different approaches and compared
to the zero point energy corrected experimental value. Self-consistent calculation using the sum of the
pseudo valence density and the soft core density (``VASP"), post-processing calculations using the
real valence density (``$\varrho_{val}$"), all-electron density with $\varrho_{ae}$ and without
$\varrho_{ae}^{no soft}$ the soft correction are given.}
\label{tab_comp_sol2}
\centerline{
\begin{ruledtabular}
\begin{tabular}{cccccc}
Solid & VASP & $\varrho_{val}$  & $\varrho_{ae}^{no soft}$ &  $\varrho_{ae}$ & Exp.(ZPEC)\\
\hline
$N_\alpha$&30&40&40&40&--\\
$\qcut$    &10&10&10&10&--\\
$\varrho_{\rm cut}$ &$\infty$&$\infty$&$\infty$& 20&--\\
\hline
Cu &3.605 & 3.607& 3.607& 3.606& 3.595\\
Rh & 3.805& 3.811& 3.813& 3.806 & 3.793\\
Pd & 3.909& 3.913& 3.912&3.912 & 3.875\\
Ag &4.101 & 4.100& 4.097& 4.098& 4.056\\
\hline
Li & 3.452& 3.454& 3.454& 3.454 &3.449 \\
Na &4.191 & 4.194& 4.185& 4.191& 4.210\\
K &5.202 &5.215 &5.213 &5.208 & 5.212\\
Rb &5.541 & 5.562& 5.548& 5.550& 5.576\\
Cs &5.945 & 5.980& 5.956&5.944 & 6.039\\
Ca & 5.465  & 5.476  & 5.471 & 5.463 & 5.553\\
Sr & 5.921 &5.937 & 5.927 &5.927 & 6.045\\
Ba & 4.906& 4.935  & 4.926 &4.920 & 4.995\\
Al & 4.036 & 4.086& 4.036 & 4.038 &4.020 \\
\hline
LiF & 4.037& 4.041& 4.039& 4.040& 3.964\\
LiCl & 5.103&5.116 & 5.109& 5.109 &5.056 \\
NaF &4.658 & 4.660& 4.656& 4.658& 4.579\\
NaCl &5.627 & 5.636& 5.625& 5.628& 5.565\\
MgO & 4.230& 4.239& 4.233& 4.234& 4.184\\
\hline
C & 3.572&3.573 &3.571 &3.571 & 3.543\\
SiC & 4.369& 4.385& 4.367& 4.367& 4.342\\
Si &5.447 &5.478 &5.458 & 5.447& 5.416\\
Ge & 5.725&5.737 & 5.728& 5.731&5.640 \\
GaAs &5.717 & 5.744& 5.724& 5.722& 5.638\\
\end{tabular}
\end{ruledtabular}}
\end{table}

\section{All electron atomization energies}

To assess the validity of our implementation and to evaluate its accuracy we calculate the atomization
energies of the solids within VASP and with the all-electron post-processing correction.
A self-consistent calculation using optB86b-vdW is done for the solid close to the energy minimum and the respective atom
or atoms in a large rectangular box (with approximately two times larger sides).
From this calculation we obtain the approximate $E_c^{nl}$. In the next step,
the all-electron density from the VASP calculation is used to evaluate $E_c^{nl,ae}$. By subtracting the solid
energies per atom from the atomic ones, we obtain the non-local contribution to the atomization energy for
these two approaches which we can compare. The results are summarized in Table~\ref{tab_at_check} where we show 
the $E_c^{nl}$ contribution to binding from VASP and from the all-electron calculations with and without
the soft-correction, along with the total atomization energy. 
The agreement is very good overall with the errors in the total atomization energies below 2\%.
The soft-correction is not calculated in VASP, however, its effect on the atomization energies is negligible.

As was shown in the case of lattice constants the all-electron density based $E_c^{nl,ae}$ strongly depends on the
underlying grid and high density regions need to be cut. The problem is less severe in the case of atomization energies
where we use exactly the same grid spacing for the solid and atomic calculations so that numerical inaccuracies
cancel out. In most cases the calculations with cut density give the same vdW contribution to the
atomization energy (to within a meV). In a few cases the contribution differs slightly (by up to 30~meV for Pd), 
and therefore we give the results calculated with the electron density cut above 100~a.u. in Table~\ref{tab_at_check}.

\begin{table}[h]
\caption[Comparison of atomization energies from VASP and all-electron calculations]
{Total atomization energy from VASP ($E_{\rm at, total}$) using optB86b-vdW and contribution of the
non-local correlation term ($E_c^{nl}$) to the atomization energies for a set of solids. The non-local correlation
has been calculated with VASP self-consistently, and using the SIESTA routine on the all-electron density with
and without the soft-correction. All data in eV. The reference atom calculations were done in a rather small cells 
with a side length approximately twice the side of the conventional unit cell of the appropriate solid
and thus they are not fully converged. The atomisation energies are therefore lower than those reported 
in Table~\ref{tab_comp_atom}.}
\label{tab_at_check}
\centerline{
\begin{ruledtabular}
\begin{tabular}{ccccc}
     &$E_{\rm at, total}$&  \multicolumn{3}{c}{$E_{\rm at}^{nl}$}\\
     &   & VASP & ae &  ae, no soft\\
\hline
Cu & 3.679&0.941 &0.933 &0.935 \\
Ag & 2.887 &1.097 &1.063 &1.065 \\
Pd & 4.160&1.309 &1.259 &1.261 \\
Rh & 6.385&1.498 &1.430 &1.433 \\
\hline
Li & 1.203&0.155 & 0.153& 0.155\\
Na &0.923 & 0.221& 0.221 &0.222\\
K &0.832 & 0.281 & 0.275&0.276\\
Rb & 0.762&0.314 & 0.301 &0.303\\
Cs & 0.743& 0.355 & 0.340&0.342\\
Ca &1.940&0.584 &0.569 &0.571  \\
Sr& 1.691& 0.638&0.616 &0.617  \\
Ba& 2.036&0.739 & 0.711& 0.713 \\
Al & 3.488& 0.799& 0.750&0.752 \\
\hline
LiF &4.410 &0.273 & 0.267& 0.268\\
LiCl & 3.489& 0.375&0.369 &0.370 \\
NaF & 3.959& 0.270 & 0.268&0.269\\
NaCl &3.239 & 0.359& 0.356& 0.357\\
MgO &5.189 & 0.533& 0.523 &0.524\\
\hline
C & 7.910& 0.772& 0.745& 0.747\\
SiC &6.669 &0.767 & 0.737& 0.738\\
Si & 4.836& 0.741& 0.712 &0.714\\
Ge & 4.003& 0.752& 0.733& 0.734 \\
GaAs & 3.442& 0.741& 0.717& 0.718\\
\end{tabular}
\end{ruledtabular}}
\end{table}

\bibliography{TS}
\end{document}